%% file: Manuscript.tex
\documentclass[conference]{IEEEtran}

\input{tex/packages.tex}

\newif\ifdoubleblind 

\renewcommand{\baselinestretch}{0.97}

\begin{document}

\input{tex/variables.tex}
\input{tex/acro.tex}

\acresetall
\input{tex/title}
\input{tex/arxiv}
\input{tex/abstract.tex}

\IEEEpeerreviewmaketitle

\input{tex/introduction.tex}
\input{tex/relatedWork.tex}

\input{tex/solution_approach.tex}

\input{tex/methodology.tex}

\input{tex/results.tex}
\input{tex/conclusion.tex}

\input{tex/acknowledgment.tex}

\bibliographystyle{IEEEtran}
\bibliography{Bibliography}

\end{document}

%% file: tex/packages.tex
\usepackage{graphicx}
\usepackage{epstopdf}
\usepackage{standalone}
\usepackage[nolist ]{acronym}
\usepackage{tcolorbox}

\usepackage{pdfpages}
\usepackage{tikz}
\usetikzlibrary{positioning}
\usepackage[bookmarks=false]{hyperref}
\usepackage{pdfcomment}
\usepackage{hologo}

\usepackage{xstring}

\usepackage[utf8]{inputenc}
\usepackage{marvosym}

\usepackage{algorithm}
\usepackage{algpseudocode}
\usepackage{tikz}

\usepackage{listings}
\definecolor{ListingBackground}{rgb}{0.97,0.97,0.97}
\usepackage{pgfplots}
\pgfplotsset{compat=newest}
\pgfplotsset{
    box plot/.style={
        /pgfplots/.cd,
        fill=blue!30,
        only marks,
        mark=-,
        mark size=0.2em,
        /pgfplots/error bars/.cd,
        y dir=plus,
        y explicit,
    },
    box plot box/.style={
        /pgfplots/error bars/draw error bar/.code 2 args={%
            \draw  ##1 -- ++(.2em,0pt) |- ##2 -- ++(-.2em,0pt) |- ##1 -- cycle;
        },
        /pgfplots/table/.cd,
        y index=2,
        y error expr={\thisrowno{3}-\thisrowno{2}},
        /pgfplots/box plot
    },
    box plot top whisker/.style={
        /pgfplots/error bars/draw error bar/.code 2 args={%
            \pgfkeysgetvalue{/pgfplots/error bars/error mark}%
            {\pgfplotserrorbarsmark}%
            \pgfkeysgetvalue{/pgfplots/error bars/error mark options}%
            {\pgfplotserrorbarsmarkopts}%
            \path ##1 -- ##2;
        },
        /pgfplots/table/.cd,
        y index=4,
        y error expr={\thisrowno{2}-\thisrowno{4}},
        /pgfplots/box plot
    },
    box plot bottom whisker/.style={
        /pgfplots/error bars/draw error bar/.code 2 args={%
            \pgfkeysgetvalue{/pgfplots/error bars/error mark}%
            {\pgfplotserrorbarsmark}%
            \pgfkeysgetvalue{/pgfplots/error bars/error mark options}%
            {\pgfplotserrorbarsmarkopts}%
            \path ##1 -- ##2;
        },
        /pgfplots/table/.cd,
        y index=5,
        y error expr={\thisrowno{3}-\thisrowno{5}},
        /pgfplots/box plot
    },
    box plot median/.style={
        /pgfplots/box plot
    },
    boxplot/every median/.style={
    	ultra thick,dashed,cyan
    }
}

\definecolor{flexicolor}{RGB}{46,49,146}
\definecolor{amaricolor}{RGB}{237,28,36}

\usepackage{xspace}

\usepackage[binary-units=true]{siunitx}
\usepackage{psfrag}
\usepackage{graphicx}
\usepackage{tabularx,booktabs}
\usepackage{multirow}
\usepackage{rotating}

\renewcommand{\baselinestretch}{1}

\usepackage{multirow}
\usepackage[cmex10]{amsmath}
\usepackage[caption=false,font=footnotesize]{subfig}

\usepackage{stfloats}

\DeclareMathOperator{\atantwo}{atan2}

%% file: tex/variables.tex
\newcommand{\paperTitle}{Template}
\newcommand{\paperAuthors}{Benjamin Sliwa and Christian Wietfeld}
\newcommand{\paperEmails}{$\{$Benjamin.Sliwa, Christian.Wietfeld$\}$@tu-dortmund.de}
\newcommand{\figurePadding}{0pt}
\newcommand{\figureTopPadding}{\figurePadding}
\newcommand{\figureBottomPadding}{\figurePadding}

\newcommand{\ben}[1]{\colorbox{yellow}{Ben: #1}}
\newcommand{\jakob}[1]{\colorbox{green}{Jakob: #1}}
\newcommand{\red}[1]{\colorbox{red}{\textbf{TODO}: #1}}

\newcommand{\dummy}[3]
{
	\begin{figure}[b!]  
		\begin{tikzpicture}
		\node[draw,minimum height=6cm,minimum width=\columnwidth]{\LARGE #1};
		\end{tikzpicture}
		\caption{#2}
		\label{#3}
	\end{figure}
}

\newcommand{\wDummy}[3]
{
	\begin{figure*}[b!]  
		\begin{tikzpicture}
		\node[draw,minimum height=6cm,minimum width=\textwidth]{\LARGE #1};
		\end{tikzpicture}
		\caption{#2}
		\label{#3}
	\end{figure*}
}

\newcommand{\basicFig}[7]
{
	\begin{figure}[#1]  	
		\vspace{#6}
		\centering		  
		\includegraphics[width=#7\columnwidth]{#2}
		\caption{#3}
		\label{#4}
		\vspace{#5}	
	\end{figure}
}
\newcommand{\fig}[4]{\basicFig{#1}{#2}{#3}{#4}{0cm}{0cm}{1}}

\newcommand{\subfig}[3]
{
	\subfloat[#3]
	{
		\includegraphics[width=#2\textwidth]{#1}
	}
	\hfill
}

\newcommand\circled[1] 
{
	\tikz[baseline=(char.base)]
	{
		\node[shape=circle,draw,inner sep=1pt] (char) {#1};
	}\xspace
}

%% file: tex/acro.tex
\begin{acronym}
	\acro{RAIK}{Regional Analysis to Infer KPIs}	
	\acro{LIMoSim}{Lightweight ICT-centric Mobility Simulation}	
	\acro{DDNS}{Data-driven Network Simulation}
	\acro{LIDAR}{Light Detection and Ranging}
	\acro{RSRP}{Reference Signal Received Power}
	\acro{MNO}{Mobile Network Operator}
	\acro{UE}{User Equipment}
	\acro{eNB}{evolved Node B}
	\acro{ANN}{Artificial Neural Network}
	\acro{5GAA}{5G Automotive Association}
	\acro{QoS}{Quality of Service}
	\acro{OSM}{OpenStreetMap}
	\acro{RSRP}{Reference Signal Received Power}
	\acro{RSS}{Received Signal Strength}
	\acro{DNN}{Deep Neural Network}
	\acro{RoI}{Region of Interest}
	\acro{RMSE}{Root Mean Square Error}
	\acro{ReLU}{Rectified Linear Unit}
\end{acronym}

%% file: tex/title.tex
\title{Deep Learning-based Signal Strength Prediction Using Geographical Images and Expert Knowledge}

\ifdoubleblind
\author{\IEEEauthorblockN{\textbf{Anonymous Authors}}
	\IEEEauthorblockA{Anonymous Institutions\\
		e-mail: Anonymous Emails}}
\else
\author{\IEEEauthorblockN{\textbf{Jakob Thrane$^1$, Benjamin Sliwa$^2$, Christian Wietfeld$^2$, and Henrik L. Christiansen$^1$}}
	\IEEEauthorblockA{
		$^1$Department of Photonics Engineering, Technical University of Denmark, 2800 Kongens Lyngby, Denmark\\
		$^2$Communication Networks Institute, TU Dortmund University, 44227 Dortmund, Germany\\
		e-mail: $^1\{$jathr, hlch$\}$@fotonik.dtu.dk $^2\{$Benjamin.Sliwa, Christian.Wietfeld$\}$@tu-dortmund.de
		}
	}
\fi

\maketitle

%% file: tex/arxiv.tex
\begin{tikzpicture}[remember picture, overlay]
\node[below=5mm of current page.north, text width=20cm,font=\sffamily\footnotesize,align=center] {Accepted for presentation in: 2020 IEEE Global Communications Conference (GLOBECOM)\vspace{0.3cm}\\\pdfcomment[color=yellow,icon=Note]{
@InProceedings\{Thrane2020deep,\\
	Author = \{Jakob Thrane and Benjamin Sliwa and Christian Wietfeld and Henrik Christiansen\},\\
	Title = \{Deep Learning-based Signal Strength Prediction Using Geographical Images and Expert Knowledge\},\\
	Booktitle = \{2020 IEEE Global Communications Conference (GLOBECOM)\},\\
	Year = \{2020\},\\
	Address = \{Taipei, Taiwan\},\\
	Month = \{Dec\}\\
\}
}};
\node[above=5mm of current page.south, text width=15cm,font=\sffamily\footnotesize] {2020~IEEE. Personal use of this material is permitted. Permission from IEEE must be obtained for all other uses, including reprinting/republishing this material for advertising or promotional purposes, collecting new collected works for resale or redistribution to servers or lists, or reuse of any copyrighted component of this work in other works.};
\end{tikzpicture}

%% file: tex/abstract.tex
\begin{abstract}
	

Methods for accurate prediction of radio signal quality parameters are crucial for optimization of mobile networks, and a necessity for future autonomous driving solutions. The power-distance relation of current empirical models struggles with describing the specific local geo-statistics that influence signal quality parameters. The use of empirical models commonly results in an over- or under-estimation of the signal quality parameters and require additional calibration studies.


In this paper, we present a novel model-aided deep learning approach for path loss prediction, which implicitly extracts radio propagation characteristics from top-view geographical images of the receiver location. In a comprehensive evaluation campaign, we apply the proposed method on an extensive real-world data set consisting of five different scenarios and more than 125.000 individual measurements. 
    

It is found that 1) the novel approach reduces the average prediction error by up to 53~\% in comparison to ray-tracing techniques, 2) A distance of $250-300$ meters spanned by the images offer the necessary level of detail, 3) Predictions with a root-mean-squared error of $\approx 6$ dB is achieved across inherently different data sources.

\end{abstract}


%% file: tex/introduction.tex
\section{Introduction}

%
%
Radio propagation modelling is a key building block for the design of wireless communication systems and represents one the foundations for network planning \cite{Taufique/etal/2017a} and network simulation \cite{Cavalcanti/etal/2018a}.
%
%
Also, the ability to forecast network quality indicators for given geographical locations -- e.g., along a vehicular trajectory -- is an enabler for anticipatory networking \cite{Bui/etal/2017a} techniques such as opportunistic data transfer \cite{Sliwa/etal/2019d}. In a recent study \cite{5GAA/2020a}, the \ac{5GAA} has pointed out the need to implement \emph{predictive \ac{QoS}} methods for enabling connected and autonomous driving.
Moreover, knowledge about the radio propagation characteristics can be exploited to infer indicators that significantly correlate to the former. An example is the prediction of the applied transmission power based on signal strength and signal quality measurements discussed in \cite{Falkenberg/etal/2018a}.

%
%
Although conventional model-based methods are suitable for comparing the behaviour of different methods under study (e.g., resource schedulers) in abstract reference scenarios, they are not able to accurately represent the radio propagation effects in complex concrete real-world environments \cite{Sliwa/Wietfeld/2019d}.

%
%
In this work, we present a hybrid approach which brings together model-based and data-driven methods for path loss prediction and for constructing radio environmental maps. We apply \emph{deep learning} -- which is known to achieve outstanding performances in the image classification domain -- on top-view \emph{images} of the receiver environment for learning geographical \emph{radio environmental prototypes}. The latter are then utilized to forecast the received power at unobserved locations.
%
%
Based on the groundwork presented in \cite{Thrane/etal/2020a} where satellite images have been exploited to infer radio quality parameters, we analyze the prediction performance based on vector images obtained from \ac{OSM} in this work. The contributions are summarized as follows:
\begin{itemize}
    \item Presentation of a novel \textbf{model-aided deep learning method} for path loss prediction based on \ac{OSM} images of the receiver environment.
    \item Evaluation of the proposed method on a large \textbf{real world data set}.
    \item The developed software is provided in an \textbf{Open Souce}\footnote{Available at \url{https://github.com/jakthra/PseudoRayTracingOSM}} way.
\end{itemize}

%
%
\fig{t}{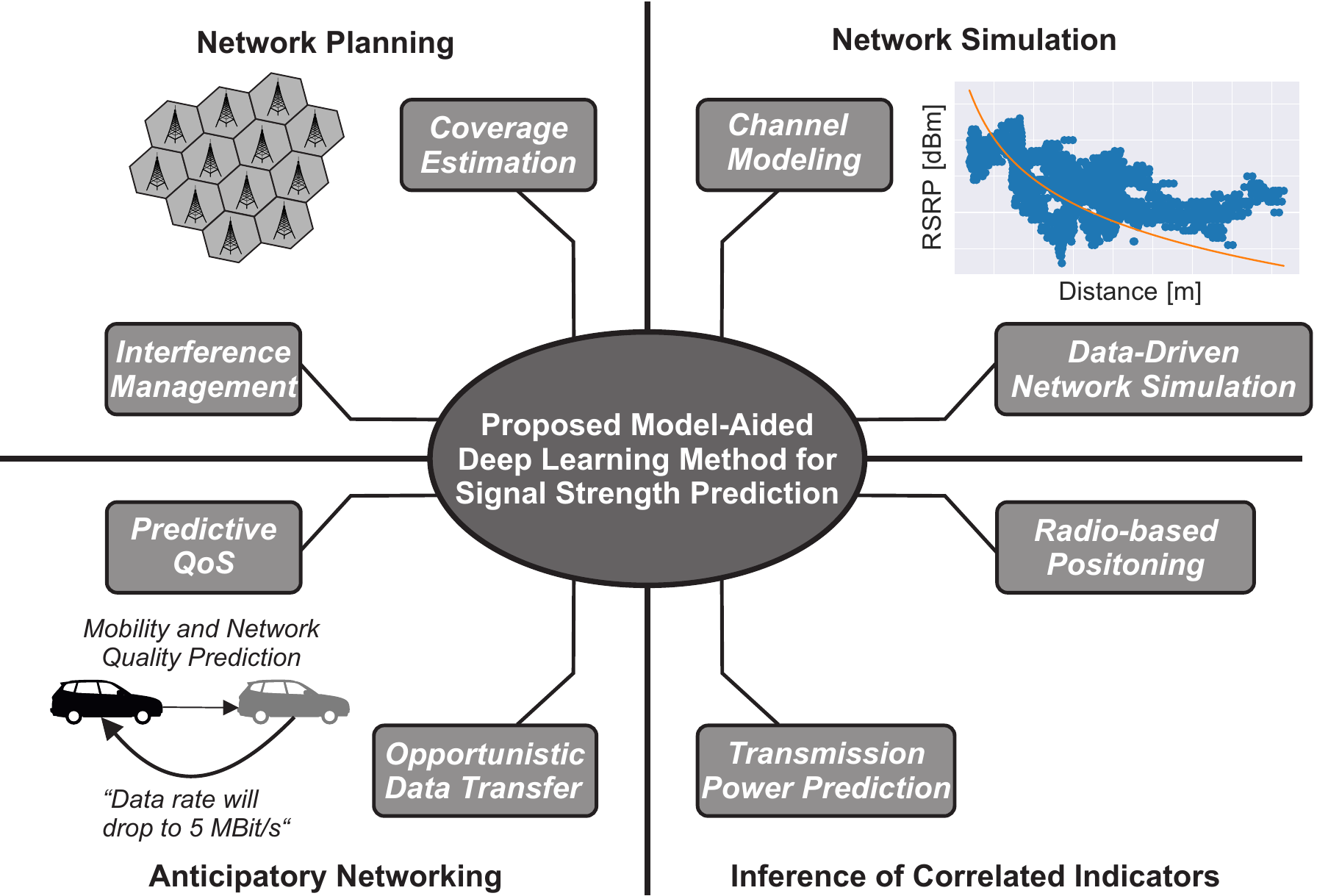}{Overview about different application scenarios for the proposed image-based signal strength prediction method.}{fig:scenario}
An overview of different target applications for the proposed method is shown in Fig.~\ref{fig:scenario}. 
%
%
The remainder of the paper is structured as follows. After discussing the related work in Sec.~\ref{sec:related_work}, we present the proposed solution approach in Sec.~\ref{sec:approach}. Afterwards, the applied methodology is introduced in Sec.~\ref{sec:methods} and finally, the results of the performance evaluation are presented and discussed in Sec.~\ref{sec:results}.

%% file: tex/relatedWork.tex
\section{Related Work} \label{sec:related_work}

%
%
\textbf{Radio propagation modelling and prediction}:
%
%
\emph{Model-based} methods represent the standard approach for considering radio propagation effects in network simulation \cite{Cavalcanti/etal/2018a} and network planning. Existing channel models -- e.g. 3GPP TR 38.901 \cite{3GPP/2019a} -- provide high computational efficiency and allow the comparison of different methods in a highly controlled environment.
However, since fading effects are typically represented as probabilistic attenuation functions and obstacles are only modelled statistically and not explicitly, these methods are mostly not able to mimic the complex radio propagation characteristics of particular real-world scenarios \cite{Sliwa/etal/2019d}.
%
%
Ray tracing methods \cite{Yun/Iskander/2015a} aim to close this gap by using detailed models of the environment to consider the dynamics of the radio link with respect to physical effects such as reflection and refraction. Although this approach is theoretically capable of providing highly-accurate representations of the radio link behaviour in concrete real-world scenarios, practical applications often lack the required amount of high-resolution data about shape and material of the obstacles. 
%
%
As an alternative to model-centric methods, data-driven approaches have emerged in recent years. Radio environmental maps \cite{Poegel/Wolf/2015a} maintain network quality information -- which is often acquired in a crowdsensing manner \cite{Raida/etal/2019a} -- based on a grid with defined cell granularity.
%
%
As discussed by \cite{Wang/etal/2019a}, incomplete measurements can be compensated by \emph{kriging}-based interpolated techniques.

%
%
\textbf{Machine learning} has gained immense popularity in the wireless networking domain as it allows to implicitly learn hidden interdependencies between measurable variables which are often too complex to bring together analytically.
%
%
A comprehensive summary of different machine learning methods and their application for various communication applications is provided by the authors of \cite{Wang/etal/2020a}.
%
%
\emph{Deep learning} \cite{Goodfellow/etal/2016a} has become one of the most popular learning methods after achieving impressive results in the image processing domain. Hereby, \emph{deep} \acp{ANN}, which consist of a high number of hidden layers, are iteratively trained to minimize a certain loss function.
%
%
A recent trend in this domain is the embedding of expert knowledge into machine learning models. Model-aided wireless artificial intelligence \cite{Zappone/etal/2019a} allows to optimize the accuracy of prediction models further and to reduce the number of required training samples.
%
%
Alongside with the convergence of machine learning and wireless communications, different authors aim to improve network quality prediction in complex scenarios by replacing the traditional mathematical radio propagation models with machine learning.
%
%
In \cite{Enami/etal/2018a}, Enami et al. propose the \ac{RAIK} framework which utilizes geographical features (e.g., the percentage of the area covered by buildings) for enhancing the prediction quality of different network performance indicators such as \ac{RSS}. Hereby, a highly-detailed environment model that consists of buildings and trees is constructed from \ac{LIDAR} information. 
%
%
A similar approach is implemented by the authors of \cite{Masood/etal/2019a}, which further considers the number of building penetrations on the direct path between transmitter and receiver. Both approaches can achieve a significant reduction of the resulting prediction error in comparison to traditional channel models.
%
%
In contrast to these existing methods, the proposed approach utilizes the raw geographical \emph{images} (instead of extracted features) of the environment between transmitter and receiver for deep learning-based \ac{RSRP} prediction. Our general assumption and motivation for this work are that similar-looking environments will likely show similar radio propagation characteristics.

%% file: tex/solution_approach.tex
\section{Machine Learning-based Signal Strength Prediction}  \label{sec:approach}

%
%
\begin{figure}[] 
    \centering
    \includegraphics[width=0.5\textwidth]{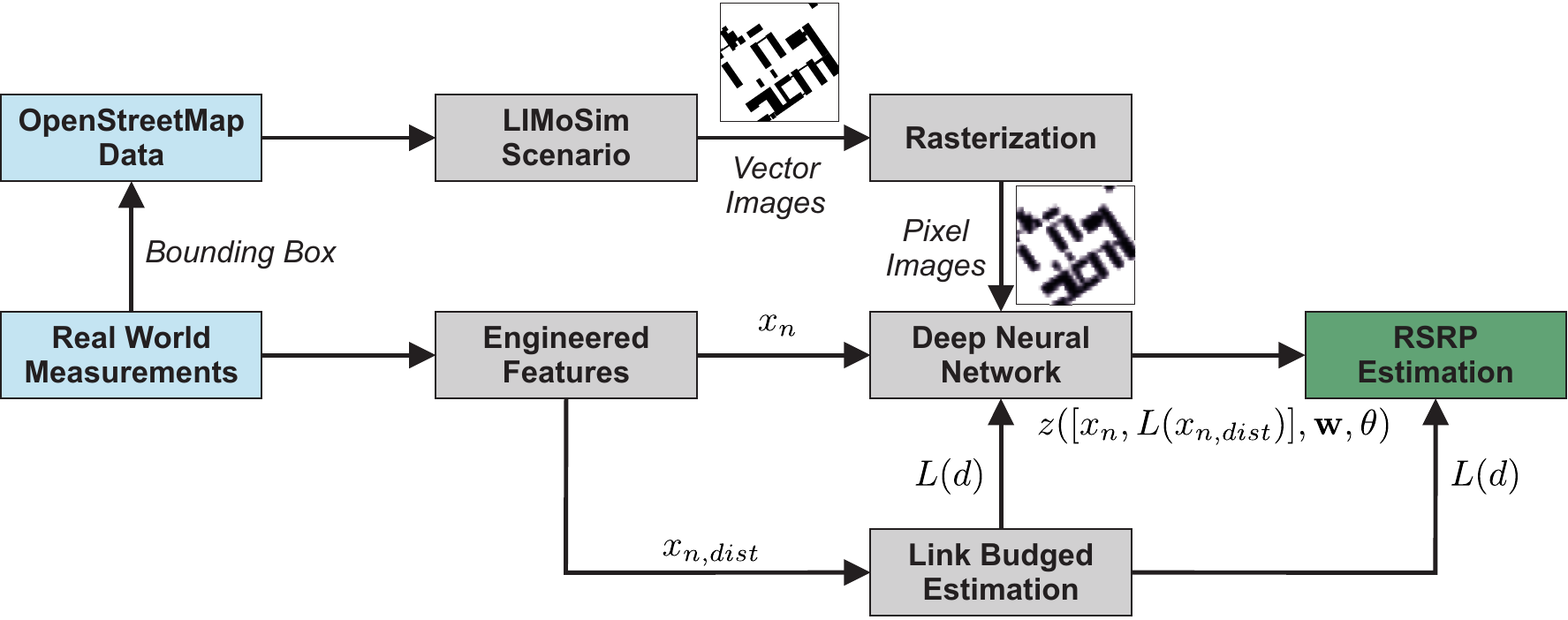}
    
    \caption{Overall system architecture model of the proposed approach.}
    \label{fig:architecture}
    
\end{figure}
%
%
The overall system architecture model for the proposed model-aided deep learning approach is shown in Fig.~\ref{fig:architecture}. The model architecture utilized for this work is identical to basic principles detailed in \cite{Thrane/etal/2020a}, however, with a few changes to 1) input parameters, i.e. features and 2) the overall model complexity. A basic path loss model is utilized to aid the learning process through a rough estimation of the link budget. The link budget consists of no learnable parameters and is based on the 3GPP UMa model \cite{3GPP/2019a}. The learning objective of the proposed model is thus to approximate the function $y(\cdot)$ that is capable of approximating the \texttt{RSRP} such that

\begin{equation}
    \text{RSRP} = y(x_{n}, \mathbf{w}, \theta) + \epsilon
\end{equation}

where $y(\cdot)$ is the model to learn, with inputs $x_n$, weights $\mathbf{w}$, and hyper-parameters $\theta$. The noise $\epsilon$ on the observed \texttt{RSRP} values is assumed Gaussian distributed, which fits well with the imposed log-normal distribution of large-scale fading. The model, $y(\cdot)$ is decomposed into a basic path loss model $L(\cdot)$ and a \ac{DNN} $z(\cdot)$.

\begin{equation}
    y(x_{n}, \mathbf{w}, \theta) = L(x_{n,dist}) + z([x_{n}, L(x_{n,dist})], \mathbf{w}, \theta)
\end{equation}

The learning objective is defined as maximizing the likelihood function through the minimization of the sum-of-squares error function between the model output, e.g. $y(x_n,...)$ and the observed values $t_n$ \cite{Thrane/etal/2020a,Goodfellow/etal/2016a}.

%
%
The observations of the \texttt{RSRP} are obtained using a drive-testing approach (see Sec.~\ref{subsec:datasets}). The features, i.e. the inputs to the models are (re)-defined as

\begin{equation}
    x_n = [v, d, \Delta_\text{lat}, \Delta_\text{lon},  f_c, \mathcal{A}]
\end{equation}

where $v$ is the vehicle's velocity, $d$ is the 3D distance between the \ac{UE} and \ac{eNB}. This is different from the work in \cite{Thrane/etal/2020a} where the raw latitude and longitude coordinates where used. These features are reduced to the differences $\Delta_\text{lat}$, and $\Delta_\text{lon}$ respectively. $f_c$ is the carrier frequency in MHz. Finally, $\mathcal{A}$ denotes an image displaying the local surroundings of the \ac{UE}.

\subsection{Generation of the Environmental Images}

%
%
For a given scenario, the corresponding \ac{OSM} map is exported based on the latitude/longitude bounding box of the scenario. The map file is then imported by the vehicular mobility simulator \ac{LIMoSim} \cite{Sliwa/etal/2019c} which converts the raw \ac{OSM} data into a simulation scenario. We utilize this uncommon approach since \ac{LIMoSim} provides an integrated engine for automatically exporting vector graphic figures of the environment. The actual generation of the environmental images that contain the \ac{RoI} is then performed in a step-wise process:
\begin{enumerate}
    %
    %
    \item \textbf{\ac{RoI} determination}: We define the \ac{RoI} as a quadratic box of width $w$ centered around the \ac{UE} location $\mathbf{P}_{\text{UE}}$ which points towards the \ac{eNB} position $\mathbf{P}_{\text{eNB}}$. 
    %
    %
    \item \textbf{\ac{RoI} rotation}: It can be assumed that different types of regions within the \ac{RoI} provide different types of information for the signal strength prediction process since they are affected by different radio propagation effects. For example, the front region $R_\text{front}$ facing the \ac{eNB} position is highly impacted by obstacle shadowing while the back region $R_\text{back}$ more likely corresponds to multipath propagation effects. For allowing the neural networks to implicitly learns these impact factors, the rotation of the images must be normalized. For a given direction vector $\mathbf{v} = \mathbf{P}_{\text{eNB}} - \mathbf{P}_{\text{UE}}$, all elements of the image are rotated around the angle $\phi = -\atantwo(\mathbf{v}.y, \mathbf{v}.x)$.
    %
    %
    \item \textbf{\ac{RoI} rasterization}: Finally, the vector image is rasterized in order to allow the further processing with the deep learning pipeline. Each pixel of the image corresponds to one input neuron of the  neural network.
\end{enumerate}
%
%
\fig{}{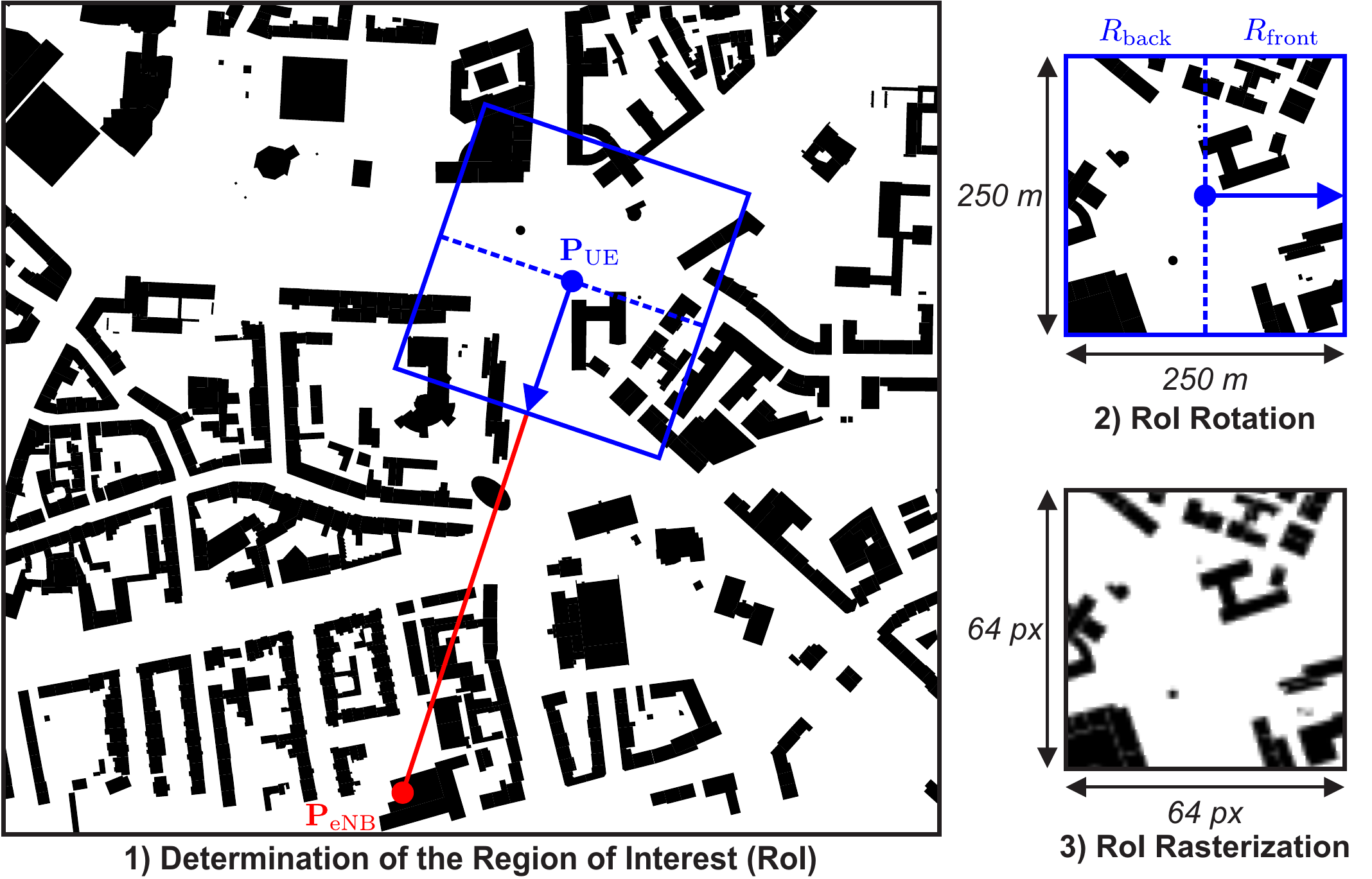}{Example of the different steps of the generation process of the environmental images based on the \texttt{GER Urban} data set. (Map data: ©OpenStreetMap contributors, CC BY-SA).}{fig:image_example}
An example which illustrates the different processing steps in a concrete scenario is shown in Fig.~\ref{fig:image_example}.

%
%
\subsection{Deep Neural Network}

The model, and function $z(\cdot)$ consists of \ac{DNN} building blocks and methodologies. The model utilizes a set of convolutional layers to process the image input, and a set of dense fully-connected layers to process the remainder of the features. In short, three sets of sub-functions are utilized within the \ac{DNN}, which transforms a set of features into valid predictions.
\begin{equation}
    z(x_n, \mathbf{w}, \mathbf{\theta}) = f[g(x_{n, A}, \mathbf{w_g}, \mathbf{\theta_g}), h(x_{n, \notin A}, \mathbf{w_h}, \mathbf{\theta_h)}, \mathbf{w_f}, \mathbf{\theta_f}]
\end{equation}
Here the function $g(\cdot)$ denotes a convolutional neural network, which consists of $4$ layers. Each layer uses a set of sequential operations. 1) 2D convolutions, 2) \ac{ReLU} activation, 3) Batch Normalization and 4) Max Pooling. The function $h(\cdot)$ is a dense neural network and is termed \emph{Feature NN}. The NN takes all features except that of the image as input. A set of fully-connected layers and a \ac{ReLU} activation function is used in the NN. The function $f(\cdot)$ is termed \emph{Output NN}, and is a set of fully-connected linear layers that adds the latent features provided by the output of $g(\cdot)$ and $h(\cdot)$. The parameters for each sub-function and layer can be found in Tab.~\ref{tab:hyper-parameters}.

%% file: tex/methodology.tex
\section{Methodology}  \label{sec:methods}

In this section, we give an overview of the evaluation scenarios, the performance metrics and the training of the Deep Learning model.

\subsection{Data Sets and Scenarios}\label{subsec:datasets}

For the following performance evaluation, we use the aggregated measurements of two large data sets which have initially been acquired in \cite{Thrane/etal/2020a} (Danish data set \texttt{DK}, single \ac{MNO}) and \cite{Sliwa/Wietfeld/2019a} (German data set \texttt{GER}, three \acp{MNO}). The resulting data set consists of more than 125.000 individual vehicular \ac{RSRP} measurements in five different scenarios with different building densities:
%
%
\begin{itemize}
    \item \textbf{DK Campus} (57586 samples): Campus area of the Technical University of Denmark
    \item \textbf{GER Campus} (8579 samples): Campus area of the TU Dortmund University
    \item \textbf{GER Urban} (11921 samples): Inner city ring of Dortmund
    \item \textbf{GER Suburban} (27152 samples): Suburban region close to campus of the TU Dortmund University
    \item \textbf{GER Highway} (20662 samples): German highway A45 with a maximum speed of up to 150~km/h
\end{itemize}

\subsection{Error Metric}

For the evaluation of the prediction performance, we focus on the \ac{RMSE} which is defined as
%
%
\begin{equation}
    \text{RMSE} = \sqrt{\frac{\sum_{i=1}^{N} \left(  \tilde{y}_{i} - y_{i} \right)^2}{N}}.
\end{equation}
with $\tilde{y}_{i}$ being the current prediction, $y_{i}$ being the current true value and $N$ being the number of samples.

\subsection{Training the model}
The proposed method is implemented using the framework PyTorch accelerated using a GTX 1080 Ti GPU. The implementation is available open-source. The well-known Adam optimizer is used for training the model through backpropagation principles. A so-called learning rate scheduler on plateau is utilized, with a patience parameter of 20 epochs, before lowering the learning rate with a factor of 10. In other words, if the test performance has not seen any improvements for 20 epochs, the learning rate is lowered. Early stopping is enforced when the learning rate is $< 1e-7$. So-called mini-batch training is utilized \cite{Goodfellow/etal/2016a}, splitting the size of the data set into smaller batches. We denote a single epoch as iterating over all mini-batches in the data set.

Each image is supplied as a \texttt{.png} image, with a fixed resolution. Data augmentation is applied to the images to improve the generalization of the model and reduce the overfitting during training. A random transformation (random rotation of $\pm 20$ degrees) is applied to the input images every epoch.

%% file: tex/results.tex
\section{Results}  \label{sec:results}

In this section, we present and discuss the results of achieved by applying the proposed method on real-world measurement data. 
%
%
\subsection{Setup Configuration, Training and Parameter Selection}

The hyper-parameters used for the resulting Deep Learning model can be found in Table \ref{tab:hyper-parameters}. The hyper-parameters found are the results of an extensive search (over 500 experiments) using Bayesian optimization techniques \cite{Goodfellow/etal/2016a} utilizing all of the available data. A significant reduction in model complexity compared to \cite{Thrane/etal/2020a} is achieved. For instance, a reduction of $\approx 170$ filters in the initial convolutional layers is achieved. The fully-connected layers are also reduced from $200$ neurons to only $32$. Furthermore, the number of convolutional layers are reduced from $6$ to $4$. 

\begin{table}[]
    \centering
    \caption{Hyper-parameters for the deep neural network model.}\label{tab:hyper-parameters}
    \begin{tabular}{@{}lll@{}}
    \toprule
    \textbf{Parameter}          & \textbf{Value}                &  \\ \midrule
    Weight decay                & 8e-4                          &  \\
    Learning rate               & 1e-3                          &  \\
    Filters                     & [32, 32, 10, 1]               &  \\
    Kernel size                 & [(5,5), (3,3), (3,3), (2,2)]  &  \\
    Max pooling                 & [2, 2, 2, 2]                  &  \\
    Feature NN layer size       & [32, 32]                      &  \\
    Output NN layer size        & [16, 16]                      & \\
    Image augmentation angle    & 20                            & \\
    Image size                  & $64~\text{px} \times 64~\text{px}$                & \\
    Batch size                  & 12                            &  \\ \bottomrule
\end{tabular}
\end{table}

%
%
\begin{figure}
    \includegraphics[width=0.5\textwidth]{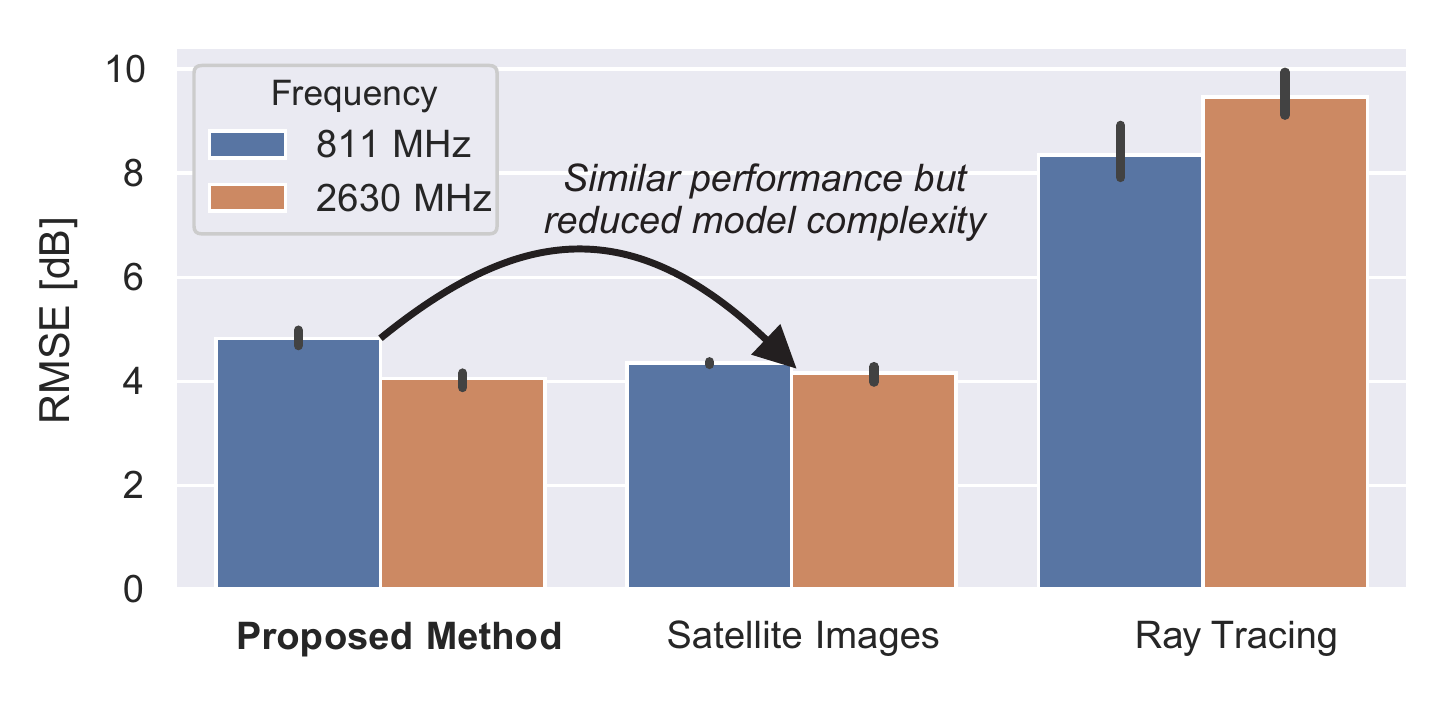}
    \caption{Performance of the proposed method compared to the use of satellite images and ray tracing for the \texttt{DK Campus} data set \cite{Thrane/etal/2020a}.}\label{fig:model_comparison_access}
\end{figure}

The method proposed in this paper is an extension of the work in \cite{Thrane/etal/2020a}. A comparison of the obtained performance can be seen in Fig. \ref{fig:model_comparison_access}. A similar level of performance is achieved, even with a significant reduction in model complexity. In comparison to ray-tracing techniques, the average \ac{RMSE} is reduced by 53\%.

%
%
\subsection{Cross-scenario Performance}

\begin{figure}
    \includegraphics[width=0.5\textwidth]{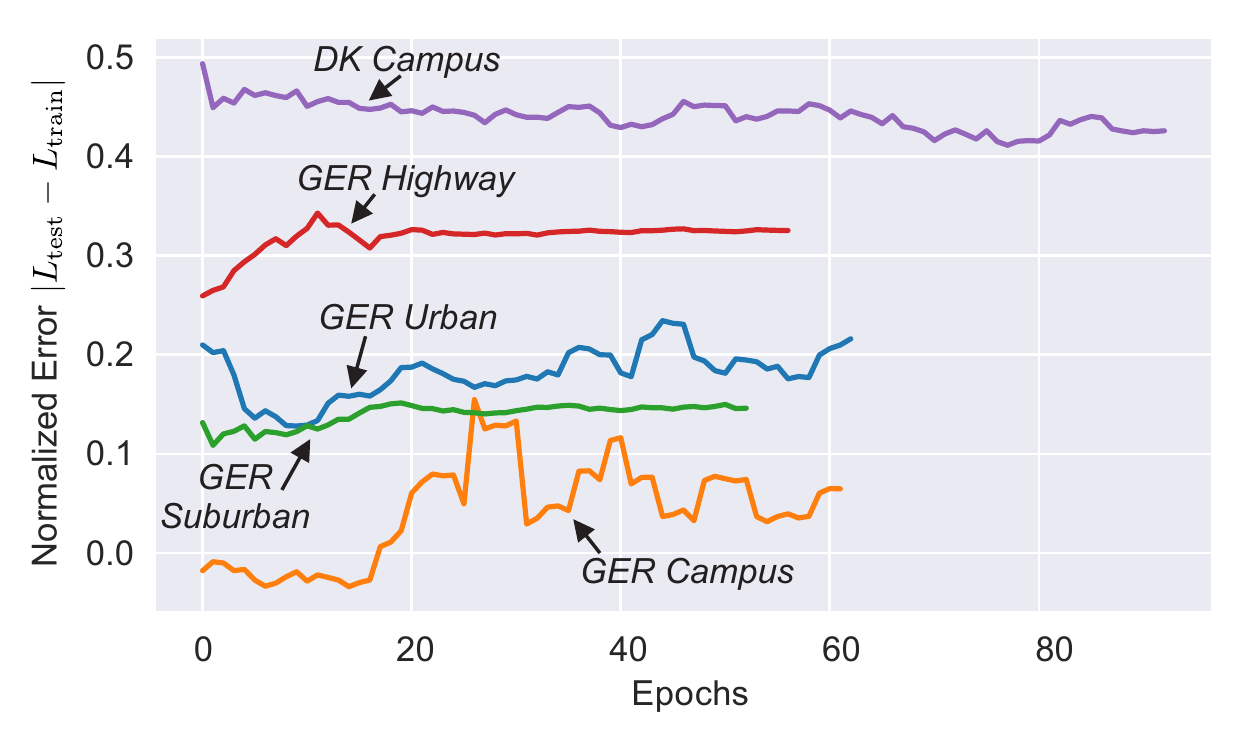}
    \caption{The difference between the training and test error over training epochs for each data subset.}\label{fig:training_test_error_crossvalidation_diff}
\end{figure}

The cross-scenario performance is evaluated in terms of RMSE. Hereby, one of the scenarios is used as the test set, while the remaining data sets compose the training set of the machine learning model. The difference between the training and test loss, utilizing the cross-scenario approach can be seen in Fig. \ref{fig:training_test_error_crossvalidation_diff}. The training and test loss is used as an indicator for the achieved generalization across data sets. If the difference is zero, the trained model is well-tuned for the problem and thus the unseen data. It can be observed that the \texttt{GER Campus} subset is well generalized if the remainder of the data subsets are used for training. This includes the use of an inherent difference data source, i.e. the \texttt{DK Campus} subset. The generalization achieved for both the \texttt{GER Suburban} and \texttt{GER Urban} scenarios are similar, however with a decrease in generalization performance compared to the \texttt{GER Campus} subset. The generalization achieved across data sources is visualized by the difference in training and test error of the \texttt{DK Campus} subset. In other words, the model is trained on a collection of subsets that all originate from the same data source. However, evaluated on a subset of data with a different data origin.

%

\begin{figure}[]
    \centering
    \includegraphics[width=0.5\textwidth]{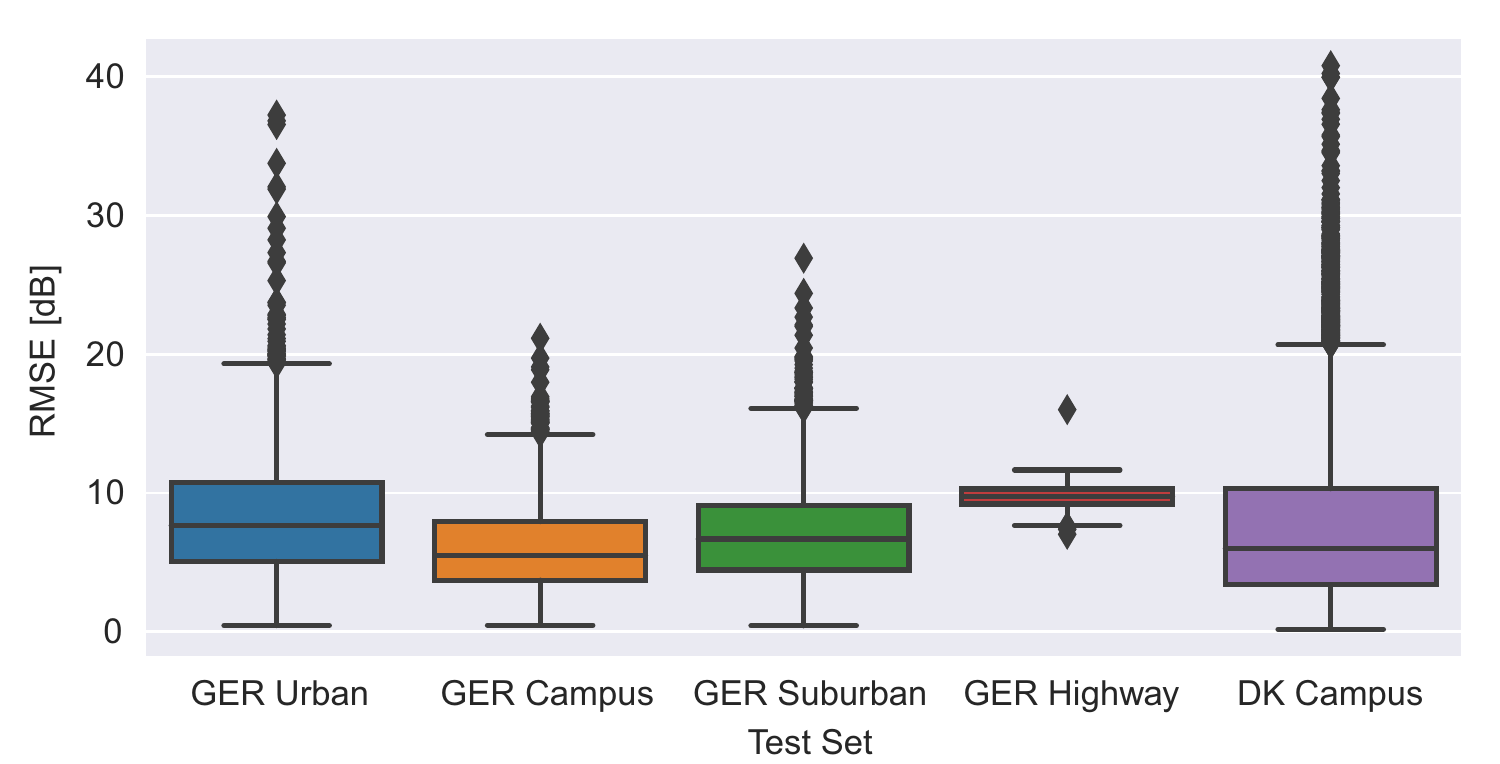}
    \caption{Cross-scenario prediction performance. For each test set, the training set is composed of the aggregation of all remaining data sets.}
    \label{fig:rmse_boxplot}
\end{figure}

The cross-scenario results for all subsets can be found in Fig. \ref{fig:rmse_boxplot}. It shows the performance of the proposed approach evaluated on each individual subset and trained on the remainder of the available subsets. In other words, the performance of the \texttt{GER Urban} scenario based on a model trained on all other subsets excluding the \texttt{GER Urban} scenario. The best generalization, also in terms of predictive performance, is achieved on the \texttt{GER Campus} subset of $6.3$ dB RMSE. Furthermore, the \texttt{GER Campus} predictions offer a significant reduction in the standard deviation  of $\sigma=3.6$ dB compared to the \texttt{DK Campus} scenario, $\sigma=6.0$ dB. The highway subset achieves the worst predictive performance of $9.7$ dB, however, does have a reduced $\sigma=1.2$ dB.

%
%

\begin{figure}[]
    \centering
    \includegraphics[width=0.5\textwidth]{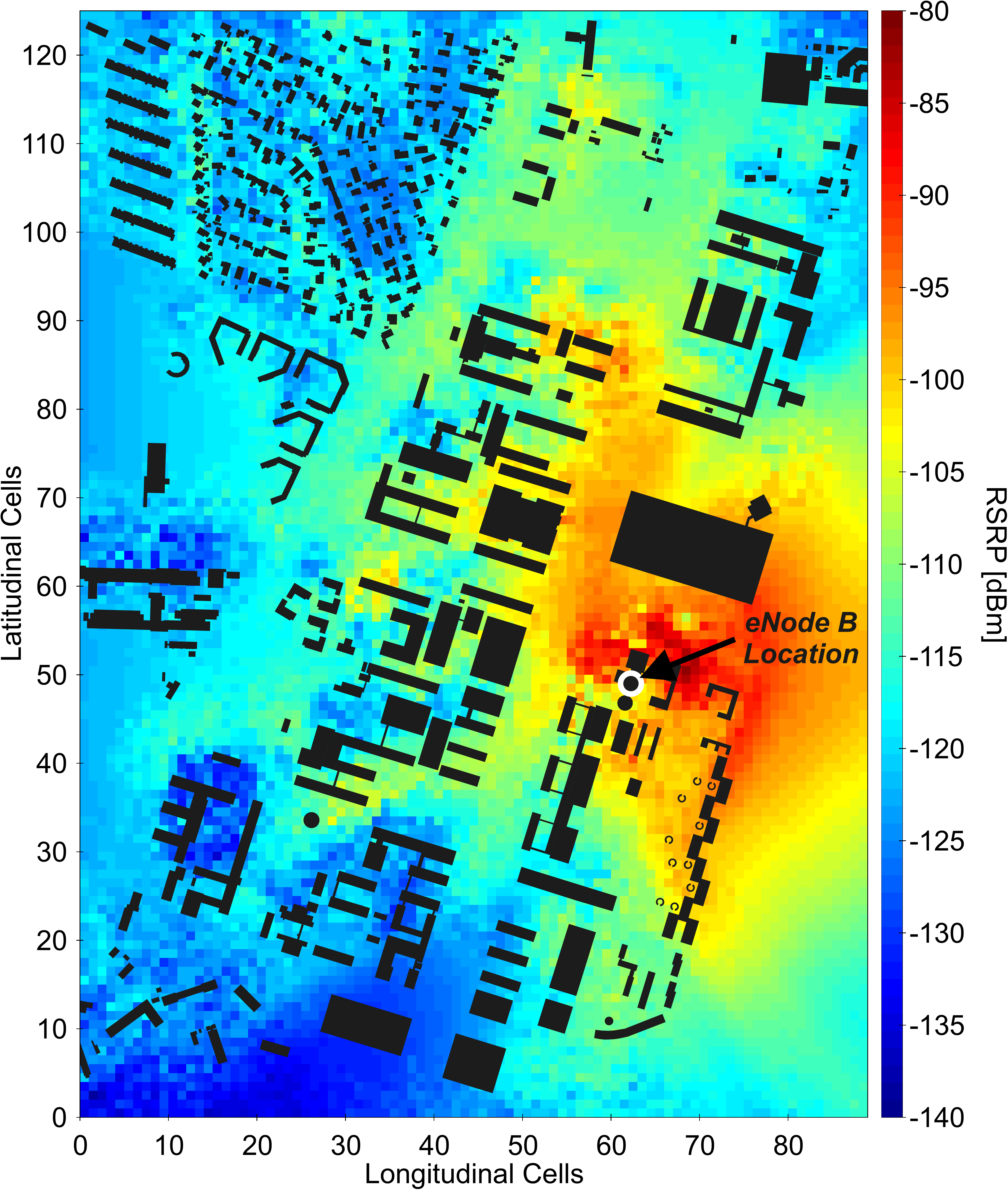}
    \caption{Resulting \ac{RSRP} radio environmental map for the \texttt{DK campus} scenario generated by the proposed method. (Map data: ©OpenStreetMap contributors, CC BY-SA).}
    \label{fig:heatmap}
\end{figure}

A heatmap of \texttt{RSRP} at 2630 MHz can be observed in Fig.~\ref{fig:heatmap}. The model is trained on all data points in the \texttt{DK Campus} scenarios. A grid of features is generated for all locations on the map. The model is then evaluated with respect to all generated features. The resulting predictions show no indication of severe overfitting. There is observed no isolated areas where the predictions are non-feasible in terms of \texttt{RSRP} magnitude. The range of predicted \texttt{RSRP} values, span from $-80$ to $-140$ dBm, with a strong increase in signal strength observed near the \ac{eNB} location.

%
%

\subsection{Comparison of Image Types}

\begin{figure}[]
    \centering
    \includegraphics[width=0.5\textwidth]{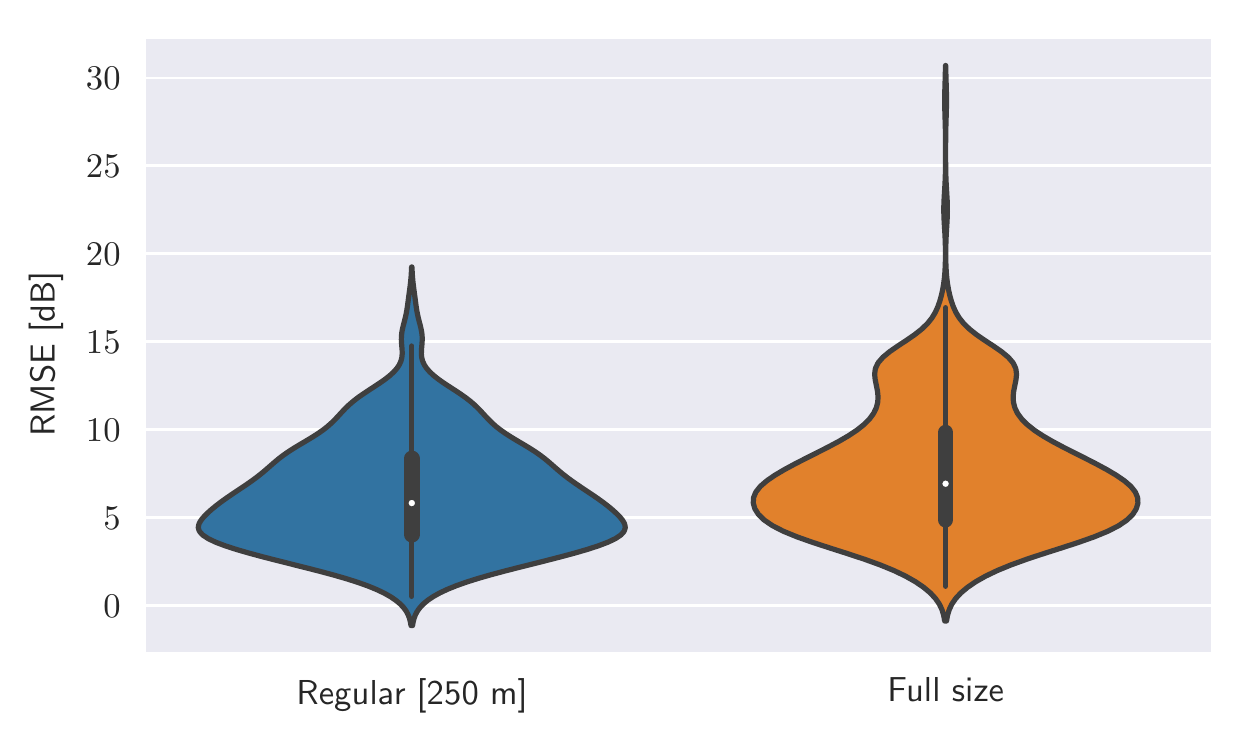}
    \caption{Comparison of different image variants for the path loss estimation process. \emph{Regular} images are receiver-centric and only contain the \ac{UE} location. For the \emph{Full size} variant, the images contain the whole transmission path between \ac{UE} and \ac{eNB}. Evaluated on the \texttt{GER Campus} subset, trained on the remainder.}
    \label{fig:rmse_violing_image_comparison}
\end{figure}

Due to the simplicity of the images and the reduced complexity of the model, significant studies of how distance and local variability is embedded in the images have been conducted. The initial intuition is that the model is learning a correction related to large-scale fading impairments using the images for deducing local variability. However, the magnitude of the large-scale fading is a complex interaction of the objects (e.g. buildings and other) in the environment. Thus, instead of providing the model with an image spanning only the local area of the measurement, a full-size image can instead be given. The performance of doing so can be observed in Fig. \ref{fig:rmse_violing_image_comparison}. The so-called full-size image is an image in which the receiver and transmitter are both localized. In other words, if the antenna separation increases the area spanned by the image increases. The performance is compared to so-called regular images, in which the area spanned is kept constant at 250 meters around the measurement position. A similar average performance is observed. The distribution of the RMSE evaluated per batch is shown as a kernel density estimation for both cases. The distribution of the RMSE using the full-size images is noticeably different. More specifically, the number and range of the outliers (i.e. predictions with a high magnitude of error) is increased. A RMSE of $6.3$ dB is observed for using the \emph{regular} images compared to $7.7$ dB for using the \emph{full size} images.

%
%

\begin{figure}
    \includegraphics[width=0.5\textwidth]{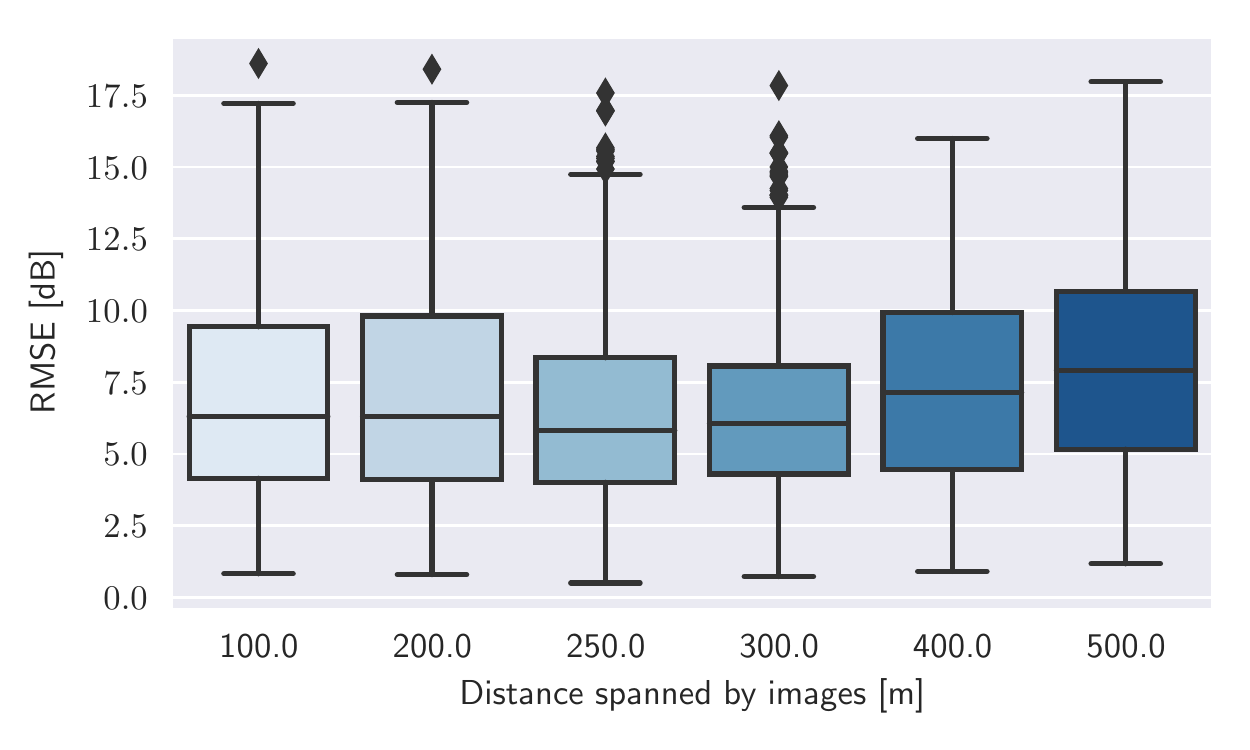}
    \caption{Comparison of different spanning distances for the receiver-centric images. Evaluated on the \texttt{GER Campus} subset, trained on the remainder.}
    \label{fig:rmse_boxplot_image_distance_comparison}
\end{figure}

A model was trained using images spanning different distances. An increase in distance results in more area covered by the image. The results of adjusting this distance can be seen in Fig. \ref{fig:rmse_boxplot_image_distance_comparison}. The best performing model was obtained using images spanning a distance of $250$ meters, and similar predictive performance was obtained using images spanning a distance of $300$ meters.

%
%

%% file: tex/conclusion.tex
\section{Conclusion}


%
%
In this paper, we presented a novel deep learning-based approach utilizing simple geographical images and expert knowledge for improving signal strength predictions in unseen locations. The approach is validated on a comprehensive collection of measurement campaigns. 
%
%
Latent features describing radio characteristics can be implicitly learned from geographical images, spanning a constant distance ($250-300$ meters across) rotated towards the transmitter location. These images contain only information on buildings in the local environment o the \ac{UE}. The proposed method is assisted by expert knowledge, ensuring optimal training conditions and improved prediction accuracy.
%
%
It is shown that the proposed approach is effective in predicting signal strength parameters in terms of predicting the \ac{RSRP} for unseen locations. Specifically, this results in an \ac{RMSE} of $\approx 6$ dB over inherently different measurement campaigns.
%
%
In future work, we want to extend the proposed method by integrating height profile information for enabling 3D-based signal strength prediction.

%% file: tex/acknowledgment.tex
\ifdoubleblind

\else
\section*{Acknowledgment}

\footnotesize
Part of the work is supported by funding provided by The Technical University of Denmark, Department of Photonics Engineering and has been supported by Deutsche Forschungsgemeinschaft (DFG) within the Collaborative Research Center SFB 876 ``Providing Information by Resource-Constrained Analysis'', project B4.
\fi